# Towards an information theoretical description of communication in brain networks

Enrico Amico[1,2], Kausar Abbas[3,4], Duy Anh Duong-Tran[4], Uttara Tipnis[3,4], Meenusree Rajapandian[4], Evgeny Chumin[6], Mario Ventresca[4], Jaroslaw Harezlak[7] and Joaquín Goñi[3,4,5,*]

* Corresponding author: jgonicor@purdue.edu
1. Institute of Bioengineering, Center for Neuroprosthetics, EPFL, Geneva, Switzerland
2. Department of Radiology and Medical Informatics, University of Geneva (UNIGE), Geneva, Switzerland
3. Purdue Institute for Integrative Neuroscience, Purdue University
4. School of Industrial Engineering, Purdue University
5. Weldon School of Biomedical Engineering, Purdue University
6. Psychological and Brain Sciences, Indiana University
7. Department of Epidemiology and Biostatistics, Indiana University

**Abstract**

Modeling communication dynamics in the brain is a key challenge in network neuroscience. We present here a framework that combines two measurements for any system where different communication processes are taking place on top of a fixed structural topology: Path Processing Score (PPS) estimates how much the brain signal has changed or has been transformed between any two brain regions (source and target); Path Broadcasting Strength (PBS) estimates the propagation of the signal through edges adjacent to the path being assessed.

We use PPS and PBS to explore communication dynamics in large-scale brain networks. We show that brain communication dynamics can be divided into three main "communication regimes" of information transfer: ***absent communication*** *(*no communication happening); ***relay communication*** *(*information is being transferred almost intact); ***transducted communication*** (the information is being transformed).

We use PBS to categorize brain regions based on the way they broadcast information. Subcortical regions are mainly **direct broadcasters** to multiple receivers; Temporal and frontal nodes mainly operate as **broadcast relay brain stations**; Visual and somato-motor cortices act as **multi-channel transducted broadcasters.**

This work paves the way towards the field of **brain network information theory** by providing a principled methodology to explore communication dynamics in large-scale brain networks.

**Introduction**

Deciphering communication dynamics in the human brain is one of the biggest open challenges in modern neuroscience (Avena-Koenigsberger et al., 2018). Communication in the brain can be measured and modeled at different spatial scales: starting from the fine-grained microscale exploration of information transfer between neuronal spikes (Quian Quiroga & Panzeri, 2009; Timme & Lapish, 2018), to inferring communication at mesoscale from electrical activity of cortical populations (Laughlin & Sejnowski, 2003; Nigam et al., 2016), up to macroscale brain networks estimated from *in vivo* magnetic resonance imaging (MRI) data; the latter being the focus of this work.

Particularly, in large-scale (MRI-based) brain networks, many hurdles have made the investigation of brain communication challenging. One issue arises from data acquisition, which outputs noisy and indirect measurements of neuronal activity (and subsequent connectivity or information transfer). Another issue is the difficulty of validating *in-silico* brain communication models, although meaningful progress has been made, see (Aerts et al., 2018; Cabral et al., 2017; Glomb et al., 2017; Ritter et al., 2013; Sanz Leon et al., 2013)). Additionally, several methodological factors such as selection of temporal scales, frequency ranges, time windows, and time-varying or lagged dependencies, can have significant impact on assessment of brain communication dynamics (Avena-Koenigsberger et al., 2018).

Nonetheless, in the last two decades, improvements in MRI hardware and development of new data acquisition sequences have allowed for application of methodologies from graph theory and dynamical systems, giving rise to the field of network neuroscience or brain connectomics (Bassett & Sporns, 2017; Fornito et al., 2016). In brain connectomics, the investigation of functional and structural connections in the human brain is modeled using tools and methods from network science (Fornito et al., 2016; Sporns, 2010). Structural connections between brain region pairs are modeled from diffusion weighted imaging data, denominated as structural connectome or structural connectivity (SC). Functional connections are modeled from functional magnetic resonance imaging data (fMRI), by measuring temporal statistical dependences between brain region pairs, usually defined as functional connectivity or functional connectome (FC). Examining human brain connectivity data offers new insights on how the integration and segregation of information in the brain relates to human behavior (Deco et al., 2015; Sporns, 2013), and how network organization may be altered in neurological diseases and disorders (Bassett & Bullmore, 2009; Fornito et al., 2015; Rosazza & Minati, 2011; Stam, 2014).

Brain connectomics has provided a proper mathematical framework upon which network neuroscientists have begun to layout several alternative models to capture and explain the complex patterns of brain communication dynamics stemming from large-scale brain networks. Pioneering work started by assessing the link between network topology and communication, from routing-based models with full knowledge of the topology of the brain network (i.e. signaling along shortest paths (de Pasquale et al., 2016; Graham, 2014)), to diffusion models "uninformed" of the topology of the network (Abdelnour et al., 2014; Raj et al., 2012). Hybrid models exploring a spectrum of communication dynamics (including search information (Goñi et al., 2013, 2014), navigation (Seguin et al., 2018), or k-shortest path ensembles (Avena-Koenigsberger et al., 2017)) have also been investigated. Recent studies have also looked into alternative network communication measures such as Markovian queuing networks (Mišić, Sporns, et al., 2014), linear transmission models of spreading dynamics (Mišić et al., 2015; Worrell et al., 2017), cooperative learning (Tipnis et al., 2018) and diffusion processes based on memory-biased random walks (Masuda et al., 2017), as well as studying asymmetries of communication in large-scale brain networks (Seguin et al., 2019).

Despite all these efforts in the development of communication models that explain human brain dynamics (Hahn et al., 2019; Joglekar et al., 2018), there is a lack of a principled theory of brain network communication, which aims to address the following question: how can one characterize the multi-fold communication regimes originating in the brain, on top of a fixed physical constrain represented by its structural connections?

As a matter of fact, human brain connectivity can be modeled by a multi-layered complex network that contains one slowly evolving structural topology (its structural connectome) and one rapidly evolving task-dependent functional architecture (its functional connectome) (Amico et al., 2019; Cole et al., 2014). In this context, there is a lack of a well-grounded mathematical framework that can associate structural and functional patterns and quantify the many facets of communication dynamics.

Here, we introduce a framework that combines two information-theoretical measurements for any system where different communication processes are taking place over a fixed structural topology. The first measurement, **Path Processing Score (PPS)**, estimates how much the brain signal has changed or transformed on a path between a source and a target brain region. A negative score is indicative of a path that is not being used for communication, a PPS around zero indicates that information is passed almost intact along a path from the source to the target, whereas a high PPS indicates that the signal has gone through considerable transformation. The second measurement, **Path Broadcasting Strength (PBS)**, estimates the propagation of the signal through the edges adjacent to the path being assessed. A low PBS indicates a *routing-based* communication along a path whereas a high PBS indicates that the communication is not specific to that path, but is also being *broadcast* or propagated through neighboring edges.

We apply these two measurements to investigate the communication dynamics in resting state and task functional MRI (fMRI) of 100 unrelated subjects from the Human Connectome Project (HCP). By assessing PPS, we show that routing communication dynamics in large-scale brain networks can be separated into three main "regimes": ***absent communication***, where no communication is happening along that path; ***relay communication***, where communication is specific to that path (i.e. unchanged or minimally changed brain signal), and ***transducted communication***, where communication is not path specific (i.e. transformed; modified brain signal). In addition to these three regimes, we show that our second metric, PBS, can quantify the spread of information transfer around the path (i.e. routing or diffused communication/broadcasting).

The information theoretical framework presented here allows for the joint assessment of structural and functional connectivity and has revealed different communication regimes across brain regions and different cognitive tasks. Furthermore, it also revealed a regional specificity in the way the brain broadcasts information, by categorizing brain regions into three main "communication modalities": **direct broadcasters** to multiple receivers (predominantly subcortical regions); **broadcast relay brain stations** (mainly limbic system); and, finally, **multi-channel transducted broadcasters** (mainly visual and somato-motor cortices).

This investigation was motivated by a need to better understand communication dynamics in large-scale brain networks, and it was partly inspired by the seminal masterpiece by Claude Shannon (Shannon, 1948). Several studies have shown functional connectivity changes across fMRI conditions (Amico et al., 2019, 2020; Cole et al., 2014; Gonzalez-Castillo & Bandettini, 2018; Mohr et al., 2016; Schultz & Cole, 2016). In other words, there is an adaptation or functional reconfiguration that occurs as subjects perform different tasks and/or switch between diferent cognitive modes (Gilson et al., 2018; Gonzalez-Castillo & Bandettini, 2018; Shine et al., 2016; Shine & Poldrack, 2018). In this paper, we further investigate how those changes can be reflected by communication regimes tracked on underlaying structural connectivity paths. This investigation was motivated by a need to better understand communication dynamics in large-scale brain networks, and it was partly inspired by the seminal masterpiece by Claude Shannon (Shannon, 1948). With this work, we introduce a new framework based on information theoretical principles to infer the basic units of information transfer in large-scale human brain networks, as well as to assess how they change and evolve between subjects or across cognitive tasks.

**Methods**

**Dataset.** The dataset of functional and structural neuroimaging data used in this work came from the Human Connectome Project (HCP, http://www.humanconnectome.org/), Release Q3. Per HCP protocol, all subjects gave written informed consent to the HCP consortium. These data contained fMRI and diffusion weighted imaging (DWI) acquisitions from 100 unrelated subjects of the HCP 900 data release (D. C. Van Essen et al., 2012; David C. Van Essen et al., 2013). All HCP scanning protocols were approved by the local Institutional Review Board at Washington University in St. Louis.

**HCP: fMRI acquisition.** We used fMRI runs from the 100 unrelated subjects of the HCP 900 subjects data release (D. C. Van Essen et al., 2012; David C. Van Essen et al., 2013). The fMRI resting-state runs (HCP filenames: rfMRI_REST1 and rfMRI_REST2) were acquired in separate sessions on two different days, with two different acquisitions (left to right or LR and right to left or RL) per day (Glasser et al., 2013; D. C. Van Essen et al., 2012; David C. Van Essen et al., 2013). The seven fMRI tasks were the following: gambling (tfMRI_GAMBLING), relational (tfMRI_RELATIONAL), social (tfMRI_SOCIAL), working memory (tfMRI_WM), motor (tfMRI_MOTOR), language (tfMRI_LANGUAGE, including both a story-listening and arithmetic task) and emotion (tfMRI_EMOTION). The working memory, gambling and motor tasks were acquired on the first day; all other tasks were acquired on the second day (Barch et al., 2013; Glasser et al., 2013). For all sessions, data from both the left-right (LR) and right-left (RL) phase-encoding runs were used to calculate connectivity matrices and averaged together. Full details on the HCP dataset have been published previously (Barch et al., 2013; Glasser et al., 2013; S. M. Smith et al., 2013).

**HCP: DWI acquisition.** We used DWI data from the same 100 unrelated subjects of the HCP 900 subjects data release (D. C. Van Essen et al., 2012; David C. Van Essen et al., 2013). The diffusion weighted (DW) acquisition protocol is covered in detail elsewhere (Glasser et al., 2013; Sotiropoulos et al., 2013). Below we mention the main characteristics. Very high-resolution acquisitions (1.25 mm isotropic) were obtained by using a Stejskal–Tanner (monopolar) (Stejskal & Tanner, 1965) diffusion-encoding scheme. Sampling in q-space was performed by including 3 shells at b=1000, 2000 and 3000 $s/mm^2$. For each shell, a corresponding 90 diffusion gradient directions and 5 b0 volumes were acquired twice, with the phase encoding (PE) direction reversed for each pair (i.e. LR and RL pairs). Directions were optimized within and across shells (i.e. staggered) to maximize angular coverage using the approach of (Caruyer et al., 2011)(http://www-sop.inria.fr/members/Emmanuel.Caruyer/q-space-sampling.php), and form a total of 270 non-collinear directions for each PE direction. Correction for echo planar acquisition and eddy-current-induced distortions in the diffusion data was based on manipulation of the acquisitions so that a given distortion manifests itself differently in different images (Andersson et al., 2003). To ensure better correspondence between the PE reversed pairs, the whole set of diffusion-weighted (DW) volumes was acquired in six separate series. These series were grouped into three pairs, and within each pair the two series contained the same DW directions but with reversed phase-encoding (i.e. a series of DW volumes with RL phase-encoding is followed by a series of volumes with LR phase-encoding).

**Brain parcellation.** We employed a cortical parcellation of 360 brain regions as recently proposed by (Glasser et al., 2016) for definition of brain network nodes. For completeness, 14 sub-cortical regions were added, as provided by the HCP release (filename "Atlas_ROI2.nii.gz"), as analogously done in previous papers (Amico et al., 2019; Amico & Goñi, 2018a, 2018b). To do so, this file was converted from NIFTI to CIFTI format by using the HCP workbench software (Glasser et al., 2013; Marcus et al., 2011) (command *-cifti-create-label* http://www.humanconnectome.org/software/connectome-workbench.html).

**HCP: fMRI preprocessing.** Data were processed following the HCP functional preprocessing guidelines (Glasser et al., 2013; S. M. Smith et al., 2013). Briefly, processing steps included: artefact removal, motion correction and registration to standard Montreal Neurological Institute space in both volumetric and grayordinate formats (i.e., where brain locations are stored as surface vertices (S. M. Smith et al., 2013)), with weak highpass temporal filtering (> 2000s full width at half maximum) applied to both formats, for slow drift removal. MELODIC ICA (Jenkinson et al., 2012) was applied to volumetric data and artifact components were subsequently identified using FSL-FIX (Salimi-Khorshidi et al., 2014). Artifacts and motion-related time courses (i.e. the 6 rigid-body parameter time-series, their backwards-looking temporal derivatives, plus all 12 resulting regressors squared) were then regressed out of both volumetric and grayordinate data (S. M. Smith et al., 2013).

For the resting-state fMRI data, we also added the following steps (Amico et al., 2019; Amico & Goñi, 2018a, 2018b): global gray matter signal was regressed out of the voxel time courses (Power et al., 2014); a bandpass first-order Butterworth filter in forward and reverse directions [0.001 Hz, 0.08 Hz] was applied (Matlab functions *butter* and *filtfilt*); voxel time courses were z-scored and then averaged per brain region, excluding outlier time points outside of 3 standard deviation from the mean, using the workbench software (Marcus et al., 2011) (workbench command *-cifti-parcellate*). For task fMRI data, we applied the same steps, with exception of a less restrictive range for the bandpass filter [0.001 Hz, 0.25 Hz].

Functional connectivity network edge weights were defined as mutual information (Cover & Thomas, 2012; Shannon, 1948) between all node pairs, calculated by uniform binning of the z-scored blood-oxygen-level-dependent (BOLD) time courses (bin widths = 0.5 standard deviation, spanning range = [-3.5 3.5] z-scored BOLD activation). This resulted in a positive symmetric connectivity matrix for each fMRI session of each subject. On top of using MI-bin equal to 0.5, we have also explored three additional z-score bin-widths (0.75, 1 and 2) within the z-score range [-3.5 to 3.5]. This binning procedure was applied to the Z-scored BOLD time-series before computing mutual information. Results shown in the Supplementary Table S1 indicate that MI pairwise measurements are stable across different bin sizes. Functional connectivity matrices from the left-right (LR) and right-left (RL) phase-encoding runs were averaged to improve signal-to-noise ratio (as done in (Finn et al., 2015)). The functional connectomes were kept in its weighted form (as measured by mutual information), hence neither thresholded nor binarized.

Finally, the resulting individual functional connectivity matrices were ordered (rows and columns) according to seven resting-state cortical networks (RSNs) as proposed by Yeo and colleagues (Yeo et al., 2011). For completeness, an 8th sub-network including the 14 HCP sub-cortical regions was added (as analogously done in recent papers (Amico et al., 2019; Amico & Goñi, 2018a, 2018b)).

**HCP: DWI preprocessing.** The HCP DWI data were processed following the MRtrix3 (Tournier, Calamante, & Connelly, 2012) guidelines (for the full documentation see http://mrtrix.readthedocs.io/en/latest/tutorials/hcp_connectome.html). The following were carried out: (1) generation of a tissue-type segmented image appropriate for anatomically constrained tractography ((R. E. Smith et al., 2012), MRtrix command *5ttgen*); (2) estimation of the multi-shell multi-tissue response function ((Christiaens et al., 2015), MRtrix command *dwi2response msmt_5tt*); (3) multi-shell, multi-tissue constrained spherical deconvolution ((Jeurissen et al., 2014), MRtrix *dwi2fod msmt_csd*); (4) generation of the initial tractogram (MRtrix command *tckgen*, 10 million streamlines, maximum tract length = 250, FA cutoff = 0.06); and (5) application of the second version of Spherical-deconvolution Informed Filtering of Tractograms (SIFT2, (R. E. Smith et al., 2015)) methodology (MRtrix command tcksift2). Both SIFT (R. E. Smith et al., 2013) and SIFT2 (R. E. Smith et al., 2015, p. 2) methods provides more biologically meaningful estimates of structural connection density. However, SIFT2 allows for a more logically direct and computationally efficient solution to the streamlines connectivity quantification problem: by determining an appropriate cross-sectional area multiplier for each streamline rather than removing streamlines altogether, biologically accurate measures of white matter fiber connectivity are obtained whilst making use of the complete streamlines reconstruction (R. E. Smith et al., 2015). SIFT2 obtained streamlines were then mapped onto the 374 chosen brain regions (see Brain parcellation atlas section for details), and the average streamline length (millimeters) was calculated for all brain regions pairs (MRtrix command *tck2connectome*). Henceforth, what we will refer to as "structural connectome" represents the physical distance (in millimeters) between brain regions pairs. Here we opted for the streamline length in this case, because we wanted to link information transfer in brain networks with the sender-channel-receiver schematics proposed in electronic communication by Claude Shannon (Shannon, 1948)Therefore, we approximated the concept of communication channels to shortest paths based on structural fiber length in brain networks.(Bullmore & Sporns, 2012)

**Mathematical foundations of communication in large-scale brain structural-functional networks**

There are two main fundamental assumptions behind the framework we are proposing here. First, in order to transfer (send or receive) information directly, two brain nodes must be *structurally connected* through white matter fibers (or streamlines, as obtained through tractography)*;* second, the amount of communication taking place between two structurally connected nodes can be estimated as the functional coupling between them, here measured as the **mutual information** (Cover & Thomas, 2012) between the corresponding BOLD time series.
In summary, we define two brain regions as "communicating" when they are structurally connected and their correspondent time series show statistical dependence, with the amount of "communication" being measured through pairwise mutual information. Please note that we are using the word communication here in Shannon's sense, i.e. we are trying to characterize the amount of information shared between sent and received fMRI BOLD signals, as measured through mutual information.

Starting from these two assumptions, we here lay the basis for an information-theoretical evaluation of communication following (structural) *shortest-paths* in human large-scale brain networks. Note however that, although this work focused on communication along shortest paths, the proposed framework can be generalized to any existing path.

**Assessment of well-behaved communication along shortest paths.** Part of the conceptualization of this framework was strongly inspired by seminal work by Claude Shannon, "*A Mathematical Theory of Communication*"; particularly one main concept stemming from that work: the concept of Data Processing Inequality (DPI, (Cover & Thomas, 2012)). In brief, the DPI Theorem states that, in a Markov chain of three random variables X, Y, Z, where X→Y→Z, then MI(X ;Y) ≥ MI(X; Z), where MI(X;Y) and MI(X;Z) denote the mutual information between X and Y and between X and Z respectively. Note that this theorem can be easily extended to chains larger than N=3 (Cover & Thomas, 2012).

In other words, processing Y cannot add new information about X. This theorem has a reasonable analogy. Think of the children's "telephone game" (Blackmore, 2000). Briefly, players form a line, and the first player comes up with a message and whispers it to the second player in line. The second player repeats the message to the third player, and so on. In those conditions, the message sent to player Z through "middle player" Y can never be more intact than the original version sent by the first player X; at most equal or worse (i.e. player Y might mishear player X and alter the message).

Inspired by the concept of DPI on a chain, we defined a novel brain network measure, the **path processing score (PPS)**. Let $\Pi_{s \to t}^{task}$ be the shortest path between a brain region source (S) and a brain region target (T) for a specific fMRI *task* (e.g. resting-state, language, etc.). We defined such shortest path as a sequence of nodes $\Omega_{s \to t} = \{S, K_1, K_2, ..., K_m, T\}$, starting at the source S and ending at the Target T, with *m* intermediate nodes in between. Let us define also $\Omega_{s* \to t} = \{K_1, K_2, ..., K_m, T\}$ and $\Omega_{s \to t*} = \{S, K_1, K_2, ..., K_m\}$ as the sequences of shortest-path nodes without the source and the target on the structural connectome (SC), respectively. The SC reference model used in this study is based on fiber length. Please note that such SC was not binarized or thresholded. The shortest path computation is entirely based on the group averaged positive weighted matrix defined by pairwise fiber lengths

Note also that $\Pi_{s \to t}^{task}$ is a **structural shortest-path** (i.e. obtained from the structural connectome), where weights $\{S \to K_1 \to K_2 \to ... \to K_m \to T\}$ along the path are substituted by the ***mutual information values*** calculated on the functional connectome, i.e. $\{MI(S;K_1), MI(K_1;K_2)...MI(K_{m-1};K_m), MI(K_m;T)\}$. Each term represents the mutual information between the fMRI time-series of brain regions along the structurally connected shortest path for a specific task.

The **Path Processing Score (PPS)** of a structural shortest-path associated with a specific functional task is then defined as:

$$PPS(\Pi_{s \to t}^{task}) = \sum_{i \in \Omega_{s^* \to t}} (MI(S; K_1) - MI(S; i))$$

In a nutshell, PPS estimates how much the signal has changed or been transformed between any source and target in the brain network. In a sense, it is a relaxation of the data processing inequality; a more qualitative measurement than the Shannon's "strict" data processing theorem. This choice is based on the idea that, in human MRI brain networks, It is extremely likely that communication between two brain regions can happen on non-shortest paths (Avena-Koenigsberger et al., 2017; Goñi et al., 2014; Tipnis et al., 2018). Therefore, a score such as PPS allows for a more flexible exploration of the communication dynamics underlying the fixed structural topology. Note that PPS is not defined for pairs of brain regions with a shortest path that consists of one edge. Also note that PPS is a non-symmetric measurement (i.e., PPS(s→t) ≠ PPS(t→s)).

As a simple example, let us assume a shortest path defined as a simple chain of 3 nodes (source S, intermediate node X, target T). PPS in this case is simply defined as: $PPS(\Pi_{s \to t}) = MI(S; X) - MI(S; T)$ . If this difference is positive, it means that the Mutual information along the path has decreased, which then can be understood as an attenuation of the signal (or increase in noise). However, if the difference is negative, that would mean that the Data Processing Inequality is not satisfied. In that case, it adds evidence that the communication between S and T might not be traveling along that path (e.g. that communication path is unlikely to be in use for regions S and T).

The evaluation of PPS for a shortest path can tell us a lot about the communication regime taking place between source region S and target region T (see Fig. 1B). For instance, a low (or close to zero) path processing score indicates that information is passed almost intact from the source to the target: hence, we are in presence of a **relay communication** regime. Conversely, a high processing load indicates that the signal has gone through considerable transformations (due to either internal or external inputs): the shortest path is then operating in a **transducted communication** regime. Finally, if the PPS is negative, it means that, despite the relaxation of the data processing inequality theorem, communication along the shortest path is **absent**; that is, the mutual information along the path increases with respect to the mutual information of the original message.

**Assessment of information broadcasting along shortest paths.** Search information (SI) quantifies the *hiddenness* of the shortest path between a source node and a target node within the network by measuring the amount of knowledge or information in bits needed to access the path (Goñi et al., 2014; Rosvall et al., 2005; Sneppen et al., 2005). The more nested the shortest path between two brain regions, the higher its SI value. Conversely, the less hidden or integrated the path, the lower its SI value.

Inspired by this concept, we defined a measure of **Path Broadcasting Strength** (PBS). Similarly to the PPS defined earlier, PBS is measured as the SI (Goñi et al., 2014) along the structural shortest path $\Pi_{s \to t}^{task}$ , but superimposing the functional weights corresponding to pairwise mutual information between the brain regions along the structural path. Hence, let $MI_{s \to t} = \{MI(S; K_1), MI(K_1; K_2), ..., MI(K_{m-1}; K_m), MI(K_m; T)\}$ be the set of mutual information values along the shortest path, and $W = \{w_S, w_{K1}, w_{K2} ... w_{Km}, w_T\}$ be the set of the nodal strength along the shortest path (again, note that the nodal strength is calculated from the mutual information values where a structural edge is present). Nodal strength of a brain region i is defined as $W_i = \sum_j MI_{ij}, \text{ for all } j \neq i$, which sums all functional connectivity values in which brain region i participates.

We can then define the Path Broadcasting Strength as:

$$PBS(\Pi_{s \to t}^{task}) = -\log_2 \left( \prod_{i \in \Omega_{s \to t^*}} \frac{MI_{s \to t}^i}{W_i} \right)$$

where $W_i$ refers to the i-th element of the ordered sequence of nodal strenghts along the path $\Pi_{s \to t}^{task}$. Analogously, $MI_{s \to t}^i$ refers to the i-th element of the $MI_{s \to t}$ ordered sequence along the path $\Pi_{s \to t}^{task}$. This equation does not take into account the bias arising from different path lengths. That is, longer shortest paths will have a tendency to yield higher PBS values. To account for this, we therefore normalize PBS:

$$\overline{PBS}(\Pi_{s \to t}^{task}) = \frac{PBS(\Pi_{s \to t}^{task})}{|\Pi_{s \to t}^{task}|}$$

where $|\Pi_{s \to t}^{task}|$ is the total sum of the shortest path length (in millimeters, in this case). Henceforth, what we will refer to as PBS is its normalized version. PBS is essentially the SI (Goñi et al., 2014) computed on the functional values superimposed on a fixed structural topology (Fig. 1C). However, conceptually the interpretation differs. In fact, measuring SI on functional edges allow us to investigate how communication propagates along shortest paths. For instance, when PBS is low, the signal is flowing primarily along the shortest path, hence communication between source and target regions takes place through a *routing mode*. Conversely, when PBS is high, the communication between a regions pair is being propagated through edges adjacent to the shortest path as well, hence operating in a *broadcasting mode*.

Therefore, we can associate to each of the two *communication regimes* defined through PPS (i.e. *relay* and *transducted*), as well as for (structurally) directly connected nodes (i.e. *direct communication*), its corresponding *communication mode (routing or broadcasting)*, for any shortest path between a brain region source S and a target T (Table 1, see also Fig. 1). Note that, by defining edge weights as mean streamline length (in millimeters), the resultant units of PBS are *bits/mm*.

| Communication regime (PPS) | Broadcasting level (PBS) | Communication Mode |
|---|---|---|
| **Direct communication** (single-edge shortest path, PPS not defined) | *Low broadcasting* → | *Single-edge routing* |
| | *High broadcasting* → | *Multi-edge routing* |
| **Absent Communication (PPS < 0)** | *No broadcasting along shortest path* | *No communication along shortest path* |

| Relay Communication (PPS ≅ 0) | Low broadcasting | → | Routing relay path |
|---|---|---|---|
| | High broadcasting | → | Broadcasting relay path |
| Transducted Communication (PPS > 0) | Low broadcasting | → | Routing transduction |
| | High broadcasting | → | Broadcasting transduction |

**Table 1.** Schematic of the different *communication regimes* based on the Path Processing Score (PPS) measurement, and their associations to the spread of information (*communication mode*) along the shortest path, as assessed through Path Broadcasting Strength (PBS).

Note that PBS is a *374x374* non-symmetric matrix, since every source-target pair in the brain network has a PBS score. Hence, based on PBS, we define two different nodal broadcasting strengths, differentiating when a brain region $k$ is a *sender* ($WBS_{sender}(k)$) or a *receiver* ($WBS_{receiver}(k)$):

$$WBS_{sender}(k) = \sum_{i}^{N} PBS_{ik}; \quad WBS_{receiver}(k) = \sum_{i}^{N} PBS_{ki}$$

where $N = 374$ (number of brain regions). Finally, we define the (symmetric) Nodal Broadcasting Strength (WBS) as the average, per brain region $k$, of both measurements:

$$WBS(k) = \frac{WBS_{sender}(k) + WBS_{receiver}(k)}{2}$$

**DMN-based model for identification of communication regimes.** We defined the boundaries of the Relay Communication regime based on the PPS distribution obtained by considering only all pairwise within DMN interactions. The Default-Mode-Network (DMN) at rest as a highly coherent integrated functional network. Hence we used DMN at rest for setting the boundaries of relay communication with respect to broken and transducted communication. Therefore, for each subject, we obtained the DMN-based shortest-paths and their corresponding PPS for resting state. Finally, the boundaries for a PPS to be considered "close to zero" or in *relay communication* were set to the [5,95] percentiles of the DMN-based distribution, specifically to the PPS range [-0.04 0.07] (see Fig. S1).

Here we used Path Processing Score (PPS) and Path Broadcasting Strength (PBS) to investigate, respectively, the communication regimes and communication modes of large-scale brain networks in 100 HCP subjects, for resting-state and seven different cognitive tasks (see **HCP: fMRI acquisition for details).** The scheme depicted in Fig. 1 provides a summary of these two information-theoretical measurements of brain communication.

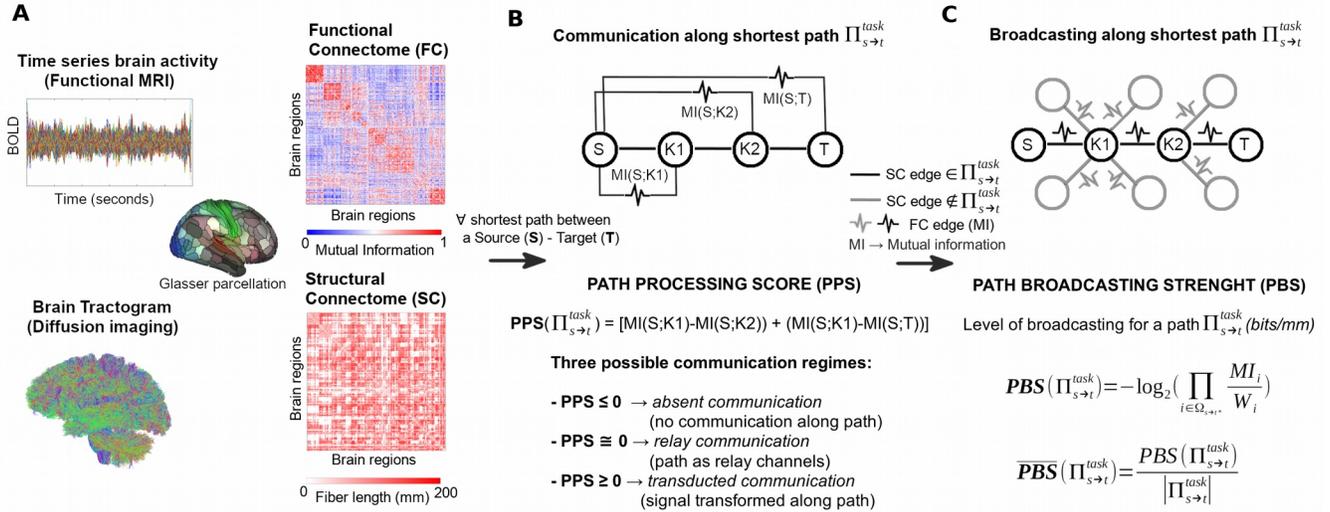

**Figure 1. Towards a mathematical theory of communication for the human connectome. A)** Functional and Structural connectomes are extracted from brain data for a multimodal brain parcellation (Glasser et al., 2016). **B)** For every shortest path between a source-target pair of brain regions, Path Processing Score (PPS) is computed and the path is assigned to its correspondent communication regime. **C)** For each communication regime, Path Broadcasting Strength (PBS) is evaluated to determine the spread of information across the shortest path.

## Results

We evaluated communication dynamics in large-scale human brain networks obtained from the MRI dataset of 100 unrelated subjects (David C. Van Essen et al., 2013), under resting-state and task conditions. Task results presented in Fig. 2A2 and 2B2 refer to the reasoning task (see Supplementary Fig. S2 for the other 6 fMRI tasks). First, we used the Path Processing Score measurement (PPS, see Methods) to characterize the shortest paths based on the three different communication regimes (*absent*, *relay*, and *transduced*; Fig. 2A1-A2). Note that the boundaries for the communication regimes were calculated from a DMN-based PPS distribution obtained from resting state data (for details see **Methods: DMN-based model for communication regimes**; Fig. S1 shows the distribution obtained with dashed vertical lines indicating 5 and 95 percentiles). Also note that even though we decided in this work to focus on the group averaged connectomes (and trials), PBS and PPS are very stable when compared across subjects and source-target pair (see Supplementary Table S2).

For each of the three different communication regimes, we stratified shortest paths into the seven functional networks as defined by (Yeo et al., 2011) (adding the subcortical set as in (Amico, Marinazzo, et al., 2017)), to investigate whether communication regimes were functional networks-specific. We observed interesting structure in the distribution of communication pathways per functional network (Fig. 2): the limbic system seems to be a hub for the relay communication regime for both task and resting-state connectomes, while the streams towards visual and DMN modules are mostly present at the transducted communication regime, for both resting-state and reasoning task (Fig. 2B1,B2, gray and blue patterns; the same applies to the other tasks, see Fig. S2). Notably, for absent paths, differential patterns emerged for the resting state and reasoning task, where absent paths predominantly appeared within network at rest and between networks during the reasoning task (Fig. 2B1, B2, red patterns). A similar pattern was observed for the other task conditions (Supplementary Fig. S2). This might be related to the tendency of going out of the optimal "routing" strategy (preferential in resting) when switching to a cognitive task.

**Figure 2 Communication regimes in large-scale brain networks. A1-A2)** Path Processing Score (PPS) on indirect pathways allows to separate brain network communication in three different regimes: absent, relay communication and transducted communication). Note that a common SC reference (group average) was used to estimate shortest-paths based on fiber-length (see more details in Methods). **B1-B2)** The percentage of paths, for the three different communication regimes, corresponding to the within and between 7 functional networks source-target pairs, as specified by (Yeo et al., 2011). An eighth sub-cortical set was added for completeness.

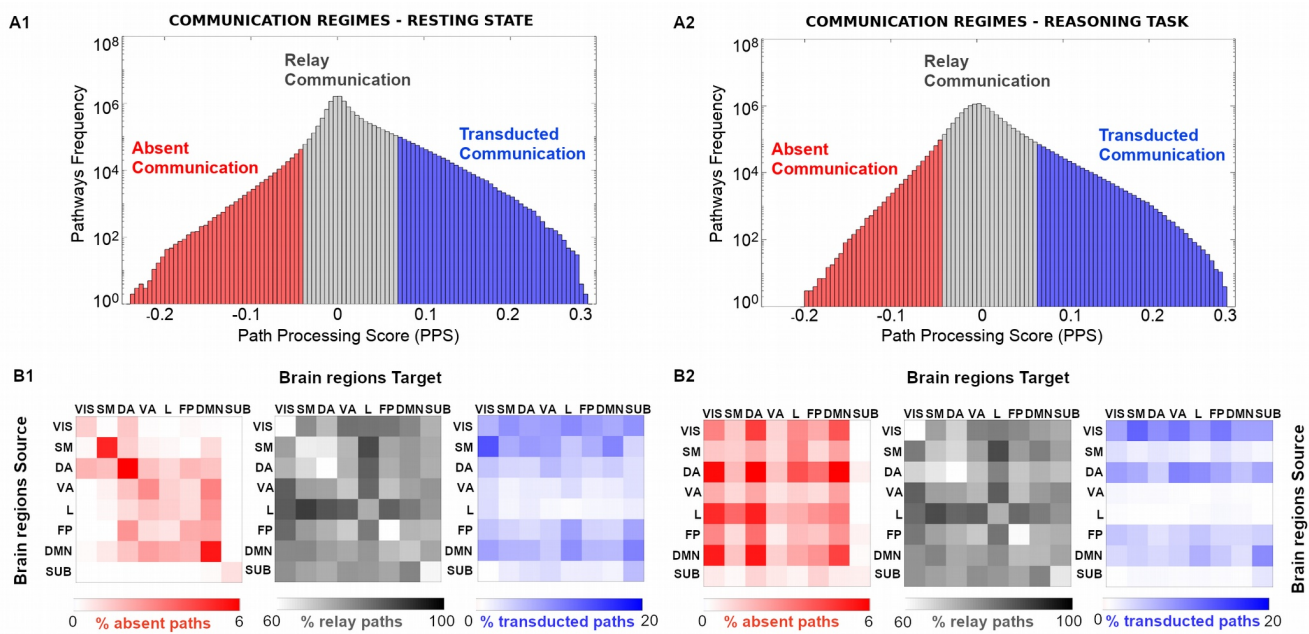

We further characterized shortest path communication regimes into two *communication modes* (routing or broadcasting) based on our second proposed metric, the Path Broadcasting Strength (PBS); see Methods and Table 1 for details). PBS quantifies the degree to which information would propagate solely along the shortest path (routing), or spread out to nodes branching from the shortest path (broadcasting). In addition to the relay and transducted regimes, which constitute paths of at least two hops, PBS was also evaluated on direct (one hop) paths (here referred to as the *Direct Communication* regime, Table 1). Notably, regional specificity emerged at each level of broadcasting (computed as Nodal Broadcasting Strength or WBS, see Methods) (Fig. 3A1-A2-A3). Specifically, within the direct communication regime, paths that involved subcortical nodes (as source/target) displayed the highest degree of broadcasting (Fig. 3A1,B1,C1; average nodal PBS of 12 *bits/mm*). For the relay communication regime, paths from/to the limbic and subcortical regions had the highest PBS (~90 *bits/mm*), operating as broadcast relay stations (Fig. 3A2,B2,C2) while, in the transducted regime pathways, visual and somatomotor cortices were the hubs of broadcasting transduction (Fig. A3,B3,C3, PBS ~ 15 *bits/mm*).

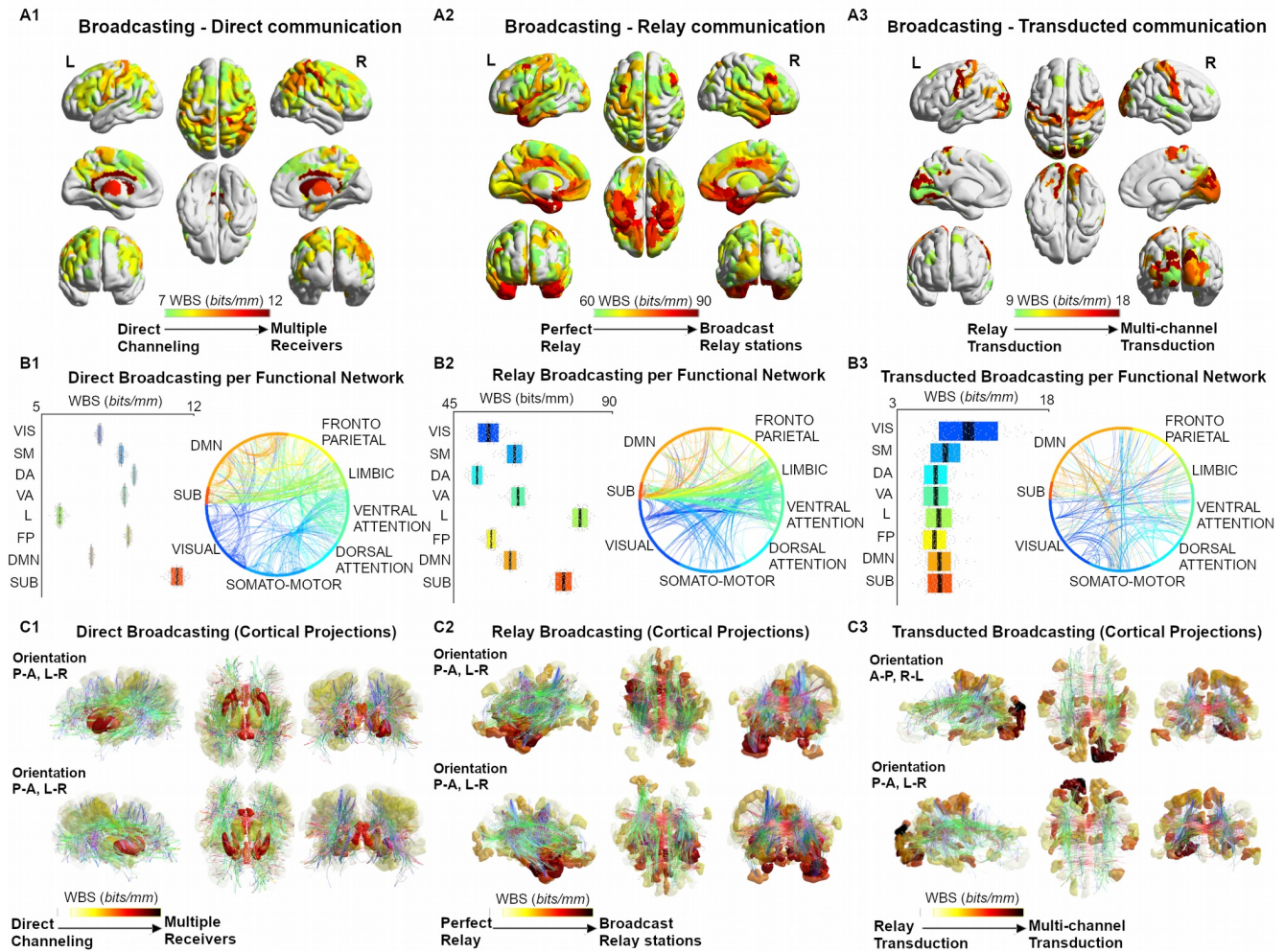

**Figure 3. Broadcasting in large-scale brain networks during rest. A1-A3)** Nodal Broadcasting strength (*WBS*, measured in *bits/mm*, see Methods for details)*,* shown for the top 100 brain regions, for the three different communication regimes (direct communication, relay communication, transducted communication). **B1-B3)** Broadcasting properties evaluated for each of the 7 functional networks specified by Yeo et al. An eighth sub-cortical community was added for completeness. **C1-C3)** The broadcasting matrices are projected onto brain renders, where tracts (color-coded by direction; Red: left-right; green: anterior- posterior; blue: superior- inferior) represent non-zero edges in the masks, and nodal strength (A1-A3) is mapped onto the cortical meshes, from low *WBS* (white; transparent) to high *WBS* (opaque; bright red).

To further investigate the top regions involved in different broadcasting scenarios, we outlined the sender-receiver broadcasting changes (using $WBS_{sender}$ and $WBS_{receiver}$, see Methods), for the top 10 brain regions in the three different communication regimes depicted in Figure 3. Overall, these regions corroborate the hypothesis of a regional specificity in the communication dynamics in large-scale human brain networks (Fig. 4). Specifically, significant source-target asymmetries were found when brain regions were broadcasting in the transducted regime (Wilcoxon rank-sum test, *p* = 8.7 *x* $10^{-05}$ for the top 10 regions, *p* = 7.9 *x*$10^{-16}$ when testing across all 374 brain nodes); tendency towards asymmetry was found for the nodal broadcasting strength in the relay regime (Wilcoxon *p = 1.8 x* $10^{-04}$ for the top 10 regions, *p* = 0.67 when testing across all 374 brain nodes); finally, no significant source-target skewness in broadcasting was found in direct communication (Wilcoxon *p = 0.66* for the top 10 regions, *p* = 0.91 when testing across all 374 brain nodes).

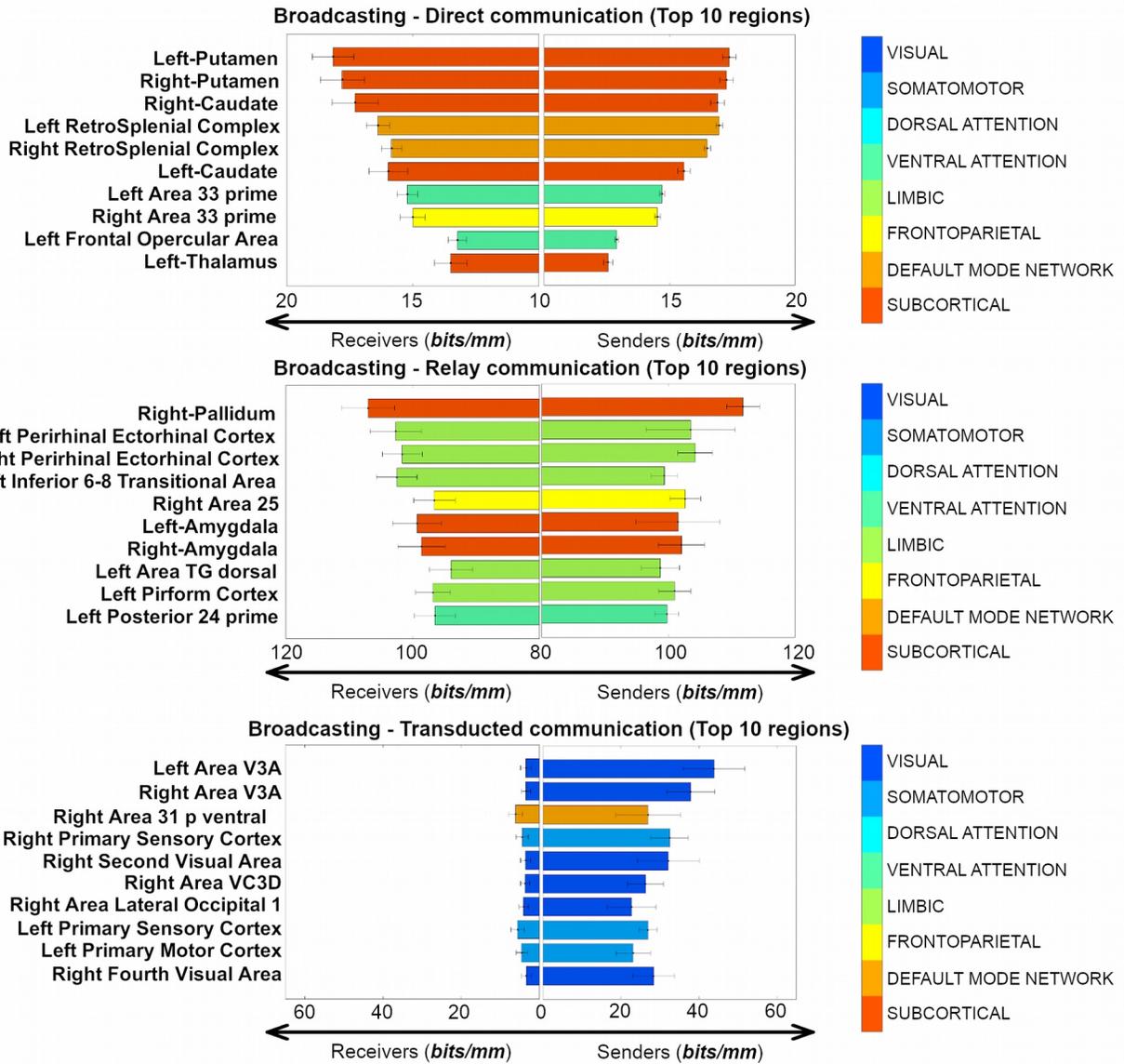

**Figure 4. Anatomical sender-receiver list of the top 10 brain regions** involved in each of the three different broadcasting regimes (direct, relay, transducted) expressed as sender and receiver nodal broadcasting strengths (WBS$_{sender}$ and WBS$_{receiver}$ respectively, both measured in *bits/mm*).

## Discussion

Understanding how the brain processes information is one of the major challenges facing the neuroscientific community in the next decade. Nonetheless, the investigation advances across different temporal and spatial scales, from neuronal population (Quian Quiroga & Panzeri, 2009) to MRI-based connectomes (Avena-Koenigsberger et al., 2018; Marinazzo et al., 2014; Mišić et al., 2015). Information theory provides a well-established mathematic framework to explore the statistical dependencies present in brain data (Wibral, Lizier, & Priesemann, 2015; Wibral, Vicente, & Lizier, 2014).

What is still lacking, in our opinion, is a theory that would allow us to investigate the information carrying capacity of a *brain network*. MRI-based connectomes can indeed be modeled as a system of multiple dynamically interacting senders and receivers (Mišić et al., 2015; Mišić, Goñi, et al., 2014; Mišić, Sporns, et al., 2014; Tipnis et al., 2018). Exploration of the presence of different communication regimes in brain networks will introduce new elements and insights in brain communication problems, such as interference and cooperation and feedback between brain regions. Extending the communication problem to a brain network level can help our understanding of how communication dynamics relate to cognitive transitions and ultimately, behavior.

In our investigation, we aim to contribute to the field by using information-theoretical tools for assessing communication dynamics in brain networks, based on their functional and structural topology. Here we introduced two information-theoretical measurements to account for communication transferred on top of a structural topology in human brain networks, specifically along the shortest paths connecting pairs of brain regions. Taking inspiration from Shannon's seminal papers on communication, we defined Path Processing Score (PPS), to serve as a quality index of how likely a shortest path is to take part in communication dynamics between a pair of regions. Using this score, we defined and explored three different regimes of communication in an MRI-based brain network: absent, relay and transducted (Fig. 2). Qualitative comparisons of communication regimes of resting-state and task (reasoning task) derived functional connectomes showed similar patterns emerging for the relay and transduction regimes, but not for absent paths (Fig. 2B). This corroborates the idea of a relationship between communication dynamics and brain functional reconfigurations (Schultz & Cole, 2016). That is, depending on the "cognitive state" in which the brain operates, communication might diffuse along many diverse paths, not necessarily the shortest.

Additionally, we define a second measurement that is complementary to PPS, termed the Path Broadcasting Strength (PBS), which is a measurement of the likelihood that communication along a path is being transferred or spread around to the neighboring nodes. Within each of the defined PPS regimes, with the addition of direct (single-edge) paths, we explored the broadcasting capacity of the resting-state connectome in the HCP dataset. Notably, we found subcortical regions (caudate, thalamus and cingulum areas) to be *broadcaster hubs* in the direct communication regimes; the limbic system (amygdala and insula cortices) to be major *broadcast relay stations*; finally, the visual and ventral cortices to be primary centers of broadcasting transduction streams (Fig. 3).

Inspired by a recent work (Seguin et al., 2019), we further explored this regional specificity by evaluating the asymmetry of broadcasting, for each communication regime, on the brain regions with highest nodal broadcasting strength. To do so, we distinguished those brain regions when being a target (receiver) or a source (sender). Interestingly, direct broadcasting showed greatest symmetry in paths originating/terminating primarily in subcortical nodes. Sender-receiver asymmetry becomes more pronounced in regions with a high broadcasting strength in relay, followed by transducted regime paths (Fig. 4).

As a matter of fact, in the case of communication of directly connected nodes, top PBS regions were those of the subcortical and attention/default networks, and showed a similar PBS magnitude when serving as either a sender or a receive node. Striatal regions are known to receive direct inputs from brainstem and cortical regions, serving to integrate information related to motor function and reward (Haber, 2016). Attention related areas (retrosplenial cortex) have been demonstrated to be involved in learning and navigation, working in concert with thalamic and hippocampal regions (Vann et al., 2009), a function that is complementary to the striatal role in motor control. Therefore, based on the findings presented here, it is likely that the direct communication regime captures activity of nodes that receive several inputs, integrate the information, and send widespread outputs to higher order cortical regions with little augmentation of the signal (Choi et al., 2012).

For the relay transduction regime (i.e. paths where the signal is not or minimally transformed on its way from source to target) half of the top ten regions belonged to the limbic network, three to subcortical, and one to each frontoparietal and ventral attention network. These nodes were primarily in the temporal lobe (perirhinal ectorhinal, amygdala, piriform) and frontal lobe (inferior 68 transitional [approximately dorsolateral prefrontal] and area 25 [subcallosal], Fig. 4), with the remaining relay nodes belonging to left posterior insula and right pallidum. In this regime, PBS values are higher when they serve as the receiver in shortest paths, as compared to being a sender. This suggests that under the relay regime, arriving information has a greater specificity to the path traveled, compared to departing (sent out) information, which has greater tendency to spread out to neighboring nodes on the path. The default mode system is commonly thought of as being active at rest, or during passive tasks, where its temporal and frontal subsystems provide information for construction and flexible use of mental simulations, respectively (Buckner et al., 2008; Yeo et al., 2011). Interpretation of our results in the context of previous work on the default mode network may hinge on the association between function and communication regime. In particular, a routing-like mode during retrieval of information from memory, and a broadcasting mode for construction and output of mental simulations (e.g. thinking about the future).

Transducted communication pathways, where signal undergoes modification on its path from source to target, showed greater broadcasting on paths where they were the source (as compared to target). Among the top regions in this regime were areas of the visual and default mode networks that were in some cases bilateral (Area V3A [visual]) or adjacent (left second and third visual areas [visual]; left area 31p ventral and area ventral 23a+b [limbic]; right fourth and eighth visual areas [visual], Fig.4). Areas of the visual network receive highly specific visual input from their receptive visual field via the lateral geniculate nuclei of the thalamus. Upon reaching the visual cortex information is propagated out to other regions via processing streams that are involved in object recognition, motion, representation in space, among others. In this regard, the information captured by PBS, from the joint structure/function connectomes, agrees with our neuroanatomical understanding of the visual system.

This study has some limitations. The impact of the brain parcellation on the definition of the communication regimes needs to be explored, as well as the choice of the soft boundaries between them (here defined on a resting-state DMN-based PPS distribution; see Methods for details); the effect of the uniform binning on the mutual information-derived connectomes should be further investigated, as well as the use of other information-based measurement of entropy between brain time series (e.g. transfer entropy or multivariate mutual information (Amico, Bodart, et al., 2017; Schreiber, 2000)). The use of resting state as a "null" condition or baseline for the tasks depends on several assumptions about neural activity during "rest" (Cole et al., 2014; Schultz & Cole, 2016). The effect of using different null conditions needs to be further investigated in terms of communication characteristics. Our study has focused on static functional connectomes as estimated by using the entire scanning length of each fMRI condition. Further studies should cover changes in communication and communication regimes within fMRI conditions by using the framework provided in this paper on dynamical functional connectivity (Hutchison et al., 2013; Preti et al., 2017; Shine et al., 2016). Finally, the choice of a reference baseline model to define the boundaries between the communication regimes (broken, relay and transducted) is somewhat arbitrary. We introduced the use of DMN at rest as the functional sub-circuit in which most of the communication should be identified as relay and hence establish relative associations of communication regimes with respect to the resulting PPS distribution (see Fig. S1 in Supplementary Material). However, other null models attending to a different rationale or also imposing different topological invariants might identify different PPS communication boundaries.

There are many possible extensions of this initial work on brain network information theory. For instance, the framework can be used on connectivity based on other modalities of brain data (as obtained via MEG, EEG, etc.), and can be extended to brain networks at different spatial scales (i.e. neuronal networks, mesoscopic brain networks). The utility of PPS and PBS for predicting behavioral, demographics and/or clinical scores should also be further investigated. Finally, one might want to consider PPS and PBS along multiple paths or path ensembles, thus not restricting to the shortest ones (Avena-Koenigsberger et al., 2017), or even select the "best communication pathway" based on PPS (or a variation of it). In the context of this framework, concepts such as interference or cooperation and feedback (Mišić et al., 2015) may additionally be included in the model. Finally, while we used fiber length,structural contribution of streamline count or a combination of the two, might be considered as well.

In conclusion, we proposed a novel methodology, rooted in information theory, to investigate communication regimes and communication modes in large-scale brain networks (i.e. **brain network information theory**). This framework sets the ground for a better characterization of brain communication regimes and how they change as subjects perform different tasks.

**Acknowledgements**

Data were provided [in part] by the Human Connectome Project, WU-Minn Consortium (Principal Investigators: David Van Essen and Kamil Ugurbil; 1U54MH091657) funded by the 16 NIH Institutes and Centers that support the NIH Blueprint for Neuroscience Research; and by the McDonnell Center for Systems Neuroscience at Washington University. JG acknowledges financial support from NIH R01EB022574, NIH R01MH108467, Indiana Alcohol Research Center P60AA07611, Purdue Discovery Park Data Science Award "Fingerprints of the Human Brain: A Data Science Perspective". EA acknowledges financial support from the SNSF Ambizione project "Fingerprinting the brain: network science to extract features of cognition, behavior and dysfunction" (grant number: PZ00P2_185716).

**Author Contributions**

E.A and J.G conceptualized the study, designed the framework, processed the MRI data and performed the analyses; all authors interpreted the results and wrote the manuscript.

**Code Availability**

The code used for computing PPS and PBS will be made available on the CONN*plexity* lab website (https://engineering.purdue.edu/ConnplexityLab).

**References**

Abdelnour, F., Voss, H. U., & Raj, A. (2014). Network diffusion accurately models the relationship between structural and functional brain connectivity networks. *NeuroImage*, *90*, 335–347. https://doi.org/10.1016/j.neuroimage.2013.12.039

Aerts, H., Schirner, M., Jeurissen, B., Van Roost, D., Achten, E., Ritter, P., & Marinazzo, D. (2018). Modeling Brain Dynamics in Brain Tumor Patients Using the Virtual Brain. *ENeuro*, *5*(3). https://doi.org/10.1523/ENEURO.0083-18.2018

Amico, E., Arenas, A., & Goñi, J. (2019). Centralized and distributed cognitive task processing in the human connectome. *Network Neuroscience*, *3*(2), 455–474. https://doi.org/10.1162/netn_a_00072

Amico, E., Bodart, O., Rosanova, M., Gosseries, O., Heine, L., Van Mierlo, P., Martial, C., Massimini, M., Marinazzo, D., & Laureys, S. (2017). Tracking Dynamic Interactions Between Structural and Functional Connectivity: A TMS/EEG-dMRI Study. *Brain Connectivity*, *7*(2), 84–97. https://doi.org/10.1089/brain.2016.0462

Amico, E., Dzemidzic, M., Oberlin, B. G., Carron, C. R., Harezlak, J., Goñi, J., & Kareken, D. A. (2020). The disengaging brain: Dynamic transitions from cognitive engagement and alcoholism risk. *NeuroImage*, *209*, 116515. https://doi.org/10.1016/j.neuroimage.2020.116515

Amico, E., & Goñi, J. (2018a). Mapping hybrid functional-structural connectivity traits in the human connectome. *Network Neuroscience (Cambridge, Mass.)*, *2*(3), 306. https://doi.org/10.1162/netn_a_00049

Amico, E., & Goñi, J. (2018b). The quest for identifiability in human functional connectomes. *Scientific Reports*, *8*(1), 1–14. https://doi.org/10.1038/s41598-018-25089-1

Amico, E., Marinazzo, D., Di Perri, C., Heine, L., Annen, J., Martial, C., Dzemidzic, M., Kirsch, M., Bonhomme, V., Laureys, S., & Goñi, J. (2017). Mapping the functional connectome traits of levels of consciousness. *NeuroImage*, *148*, 201–211. https://doi.org/10.1016/j.neuroimage.2017.01.020

Andersson, J. L. R., Skare, S., & Ashburner, J. (2003). How to correct susceptibility distortions in spin-echo echo-planar images: Application to diffusion tensor imaging. *NeuroImage*, *20*(2), 870–888. https://doi.org/10.1016/S1053-8119(03)00336-7

Avena-Koenigsberger, A., Mišić, B., Hawkins, R. X. D., Griffa, A., Hagmann, P., Goñi, J., & Sporns, O. (2017). Path ensembles and a tradeoff between communication efficiency and resilience in the human connectome. *Brain Structure and Function*, *222*(1), 603–618. https://doi.org/10.1007/s00429-016-1238-5

Avena-Koenigsberger, A., Misic, B., & Sporns, O. (2018). Communication dynamics in complex brain networks. *Nature Reviews Neuroscience*, *19*(1), 17–33. https://doi.org/10.1038/nrn.2017.149

Barch, D. M., Burgess, G. C., Harms, M. P., Petersen, S. E., Schlaggar, B. L., Corbetta, M., Glasser, M. F., Curtiss, S., Dixit, S., Feldt, C., Nolan, D., Bryant, E., Hartley, T., Footer, O., Bjork, J. M., Poldrack, R., Smith, S., Johansen-Berg, H., Snyder, A. Z., & Van Essen, D. C. (2013). Function in the human connectome: Task-fMRI and individual differences in behavior. *NeuroImage*, *80*, 169–189. https://doi.org/10.1016/j.neuroimage.2013.05.033

Bassett, D. S., & Bullmore, E. T. (2009). Human brain networks in health and disease. *Current Opinion in Neurology*, *22*(4), 340–347. https://doi.org/10.1097/WCO.0b013e32832d93dd

Bassett, D. S., & Sporns, O. (2017). Network neuroscience. *Nature Neuroscience*, *20*(3), 353–364. https://doi.org/10.1038/nn.4502

Buckner, R. L., Andrews Hanna, J. R., & Schacter, D. L. (2008). The Brain's Default Network. *Annals of the New York Academy of Sciences*, *1124*(1), 1–38. https://doi.org/10.1196/annals.1440.011

Bullmore, E., & Sporns, O. (2012). The economy of brain network organization. *Nature Reviews Neuroscience*, *13*(5), 336–349. https://doi.org/10.1038/nrn3214

Cabral, J., Kringelbach, M. L., & Deco, G. (2017). Functional connectivity dynamically evolves on multiple time-scales over a static structural connectome: Models and mechanisms. *NeuroImage*, *160*, 84–96. https://doi.org/10.1016/j.neuroimage.2017.03.045

Caruyer, E., Cheng, J., Lenglet, C., Sapiro, G., Jiang, T., & Deriche, R. (2011, September 22). *Optimal Design of Multiple Q-shells experiments for Diffusion MRI*. MICCAI Workshop on Computational Diffusion MRI - CDMRI'11. https://hal.inria.fr/inria-00617663/document

Choi, E. Y., Yeo, B. T. T., & Buckner, R. L. (2012). The organization of the human striatum estimated by intrinsic functional connectivity. *Journal of Neurophysiology*, *108*(8), 2242–2263. https://doi.org/10.1152/jn.00270.2012

Christiaens, D., Reisert, M., Dhollander, T., Sunaert, S., Suetens, P., & Maes, F. (2015). Global tractography of multi-shell diffusion-weighted imaging data using a multi-tissue model. *NeuroImage*, *123*, 89–101. https://doi.org/10.1016/j.neuroimage.2015.08.008

Cole, M. W., Bassett, D. S., Power, J. D., Braver, T. S., & Petersen, S. E. (2014). Intrinsic and Task-Evoked Network Architectures of the Human Brain. *Neuron*, *83*(1), 238–251. https://doi.org/10.1016/j.neuron.2014.05.014

Cover, T. M., & Thomas, J. A. (2012). *Elements of Information Theory*. John Wiley & Sons.

de Pasquale, F., Della Penna, S., Sporns, O., Romani, G. L., & Corbetta, M. (2016). A Dynamic Core Network and Global Efficiency in the Resting Human Brain. *Cerebral Cortex*, *26*(10), 4015–4033. https://doi.org/10.1093/cercor/bhv185

Deco, G., Tononi, G., Boly, M., & Kringelbach, M. L. (2015). Rethinking segregation and integration: Contributions of whole-brain modelling. *Nature Reviews Neuroscience*, *16*(7), 430–439. https://doi.org/10.1038/nrn3963

Finn, E. S., Shen, X., Scheinost, D., Rosenberg, M. D., Huang, J., Chun, M. M., Papademetris, X., & Constable, R. T. (2015). Functional connectome fingerprinting: Identifying individuals using patterns of brain connectivity. *Nature Neuroscience*, *18*(11), 1664–1671. https://doi.org/10.1038/nn.4135

Fornito, A., Zalesky, A., & Breakspear, M. (2015). The connectomics of brain disorders. *Nature Reviews Neuroscience*, *16*(3), 159–172. https://doi.org/10.1038/nrn3901

Fornito, A., Zalesky, A., & Bullmore, E. (2016). *Fundamentals of Brain Network Analysis*. Academic Press.

Gilson, M., Deco, G., Friston, K. J., Hagmann, P., Mantini, D., Betti, V., Romani, G. L., & Corbetta, M. (2018). Effective connectivity inferred from fMRI transition dynamics during movie viewing points to a balanced reconfiguration of cortical interactions. *NeuroImage*, *180*, 534–546. https://doi.org/10.1016/j.neuroimage.2017.09.061

Glasser, M. F., Coalson, T. S., Robinson, E. C., Hacker, C. D., Harwell, J., Yacoub, E., Ugurbil, K., Andersson, J., Beckmann, C. F., Jenkinson, M., Smith, S. M., & Van Essen, D. C. (2016). A multi-modal parcellation of human cerebral cortex. *Nature*, *536*(7615), 171–178. https://doi.org/10.1038/nature18933

Glasser, M. F., Sotiropoulos, S. N., Wilson, J. A., Coalson, T. S., Fischl, B., Andersson, J. L., Xu, J., Jbabdi, S., Webster, M., Polimeni, J. R., Van Essen, D. C., & Jenkinson, M. (2013). The minimal preprocessing pipelines for the Human Connectome Project. *NeuroImage*, *80*, 105–124. https://doi.org/10.1016/j.neuroimage.2013.04.127

Glomb, K., Ponce-Alvarez, A., Gilson, M., Ritter, P., & Deco, G. (2017). Resting state networks in empirical and simulated dynamic functional connectivity. *NeuroImage*, *159*, 388–402. https://doi.org/10.1016/j.neuroimage.2017.07.065

Goñi, J., Avena-Koenigsberger, A., Mendizabal, N. V. de, Heuvel, M. P. van den, Betzel, R. F., & Sporns, O. (2013). Exploring the Morphospace of Communication Efficiency in Complex Networks. *PLOS ONE*, *8*(3), e58070. https://doi.org/10.1371/journal.pone.0058070

Goñi, J., Heuvel, M. P. van den, Avena-Koenigsberger, A., Mendizabal, N. V. de, Betzel, R. F., Griffa, A., Hagmann, P., Corominas-Murtra, B., Thiran, J.-P., & Sporns, O. (2014). Resting-brain functional connectivity predicted by analytic measures of network communication. *Proceedings of the National Academy of Sciences*, *111*(2), 833–838. https://doi.org/10.1073/pnas.1315529111

Gonzalez-Castillo, J., & Bandettini, P. A. (2018). Task-based dynamic functional connectivity: Recent findings and open questions. *NeuroImage*, *180*, 526–533. https://doi.org/10.1016/j.neuroimage.2017.08.006

Graham, D. J. (2014). Routing in the brain. *Frontiers in Computational Neuroscience*, *8*. https://doi.org/10.3389/fncom.2014.00044

Haber, S. N. (2016). Corticostriatal circuitry. *Dialogues in Clinical Neuroscience*, *18*(1), 7–21.

Hahn, G., Ponce-Alvarez, A., Deco, G., Aertsen, A., & Kumar, A. (2019). Portraits of communication in neuronal networks. *Nature Reviews Neuroscience*, *20*(2), 117–127. https://doi.org/10.1038/s41583-018-0094-0

Hutchison, R. M., Womelsdorf, T., Allen, E. A., Bandettini, P. A., Calhoun, V. D., Corbetta, M., Della Penna, S., Duyn, J. H., Glover, G. H., Gonzalez-Castillo, J., Handwerker, D. A., Keilholz, S., Kiviniemi, V., Leopold, D. A., de Pasquale, F., Sporns, O., Walter, M., & Chang, C. (2013). Dynamic functional connectivity: Promise, issues, and interpretations. *NeuroImage*, *80*, 360–378. https://doi.org/10.1016/j.neuroimage.2013.05.079

Jenkinson, M., Beckmann, C. F., Behrens, T. E. J., Woolrich, M. W., & Smith, S. M. (2012). FSL. *NeuroImage*, *62*(2), 782–790. https://doi.org/10.1016/j.neuroimage.2011.09.015

Jeurissen, B., Tournier, J.-D., Dhollander, T., Connelly, A., & Sijbers, J. (2014). Multi-tissue constrained spherical deconvolution for improved analysis of multi-shell diffusion MRI data. *NeuroImage*, *103*, 411–426. https://doi.org/10.1016/j.neuroimage.2014.07.061

Joglekar, M. R., Mejias, J. F., Yang, G. R., & Wang, X.-J. (2018). Inter-areal Balanced Amplification Enhances Signal Propagation in a Large-Scale Circuit Model of the Primate Cortex. *Neuron*, *98*(1), 222-234.e8. https://doi.org/10.1016/j.neuron.2018.02.031

Laughlin, S. B., & Sejnowski, T. J. (2003). Communication in Neuronal Networks. *Science*, *301*(5641), 1870–1874. https://doi.org/10.1126/science.1089662

Marcus, D., Harwell, J., Olsen, T., Hodge, M., Glasser, M., Prior, F., Jenkinson, M., Laumann, T., Curtiss, S., & Van Essen, D. (2011). Informatics and Data Mining Tools and Strategies for the Human Connectome Project. *Frontiers in Neuroinformatics*, *5*. https://doi.org/10.3389/fninf.2011.00004

Marinazzo, D., Pellicoro, M., Wu, G., Angelini, L., Cortés, J. M., & Stramaglia, S. (2014). Information Transfer and Criticality in the Ising Model on the Human Connectome. *PLOS ONE*, *9*(4), e93616. https://doi.org/10.1371/journal.pone.0093616

Masuda, N., Porter, M. A., & Lambiotte, R. (2017). Random walks and diffusion on networks. *Physics Reports*, *716–717*, 1–58. https://doi.org/10.1016/j.physrep.2017.07.007

Mišić, B., Betzel, R. F., Nematzadeh, A., Goñi, J., Griffa, A., Hagmann, P., Flammini, A., Ahn, Y.-Y., & Sporns, O. (2015). Cooperative and Competitive Spreading Dynamics on the Human Connectome. *Neuron*, *86*(6), 1518–1529. https://doi.org/10.1016/j.neuron.2015.05.035

Mišić, B., Goñi, J., Betzel, R. F., Sporns, O., & McIntosh, A. R. (2014). A Network Convergence Zone in the Hippocampus. *PLOS Computational Biology*, *10*(12), e1003982. https://doi.org/10.1371/journal.pcbi.1003982

Mišić, B., Sporns, O., & McIntosh, A. R. (2014). Communication Efficiency and Congestion of Signal Traffic in Large-Scale Brain Networks. *PLOS Computational Biology*, *10*(1), e1003427. https://doi.org/10.1371/journal.pcbi.1003427

Mohr, H., Wolfensteller, U., Betzel, R. F., Mišić, B., Sporns, O., Richiardi, J., & Ruge, H. (2016). Integration and segregation of large-scale brain networks during short-term task automatization. *Nature Communications*, *7*(1), 13217. https://doi.org/10.1038/ncomms13217

Nigam, S., Shimono, M., Ito, S., Yeh, F.-C., Timme, N., Myroshnychenko, M., Lapish, C. C., Tosi, Z., Hottowy, P., Smith, W. C., Masmanidis, S. C., Litke, A. M., Sporns, O., & Beggs, J. M. (2016).

Rich-Club Organization in Effective Connectivity among Cortical Neurons. *Journal of Neuroscience*, *36*(3), 670–684. https://doi.org/10.1523/JNEUROSCI.2177-15.2016

Power, J. D., Mitra, A., Laumann, T. O., Snyder, A. Z., Schlaggar, B. L., & Petersen, S. E. (2014). Methods to detect, characterize, and remove motion artifact in resting state fMRI. *NeuroImage*, *84*, 320–341. https://doi.org/10.1016/j.neuroimage.2013.08.048

Preti, M. G., Bolton, T. A., & Van De Ville, D. (2017). The dynamic functional connectome: State-of-the-art and perspectives. *NeuroImage*, *160*, 41–54. https://doi.org/10.1016/j.neuroimage.2016.12.061

Quian Quiroga, R., & Panzeri, S. (2009). Extracting information from neuronal populations: Information theory and decoding approaches. *Nature Reviews Neuroscience*, *10*(3), 173–185. https://doi.org/10.1038/nrn2578

Raj, A., Kuceyeski, A., & Weiner, M. (2012). A Network Diffusion Model of Disease Progression in Dementia. *Neuron*, *73*(6), 1204–1215. https://doi.org/10.1016/j.neuron.2011.12.040

Ritter, P., Schirner, M., McIntosh, A. R., & Jirsa, V. K. (2013). The Virtual Brain Integrates Computational Modeling and Multimodal Neuroimaging. *Brain Connectivity*, *3*(2), 121–145. https://doi.org/10.1089/brain.2012.0120

Rosazza, C., & Minati, L. (2011). Resting-state brain networks: Literature review and clinical applications. *Neurological Sciences: Official Journal of the Italian Neurological Society and of the Italian Society of Clinical Neurophysiology*, *32*(5), 773–785. https://doi.org/10.1007/s10072-011-0636-y

Rosvall, M., Trusina, A., Minnhagen, P., & Sneppen, K. (2005). Networks and Cities: An Information Perspective. *Physical Review Letters*, *94*(2), 028701. https://doi.org/10.1103/PhysRevLett.94.028701

Salimi-Khorshidi, G., Douaud, G., Beckmann, C. F., Glasser, M. F., Griffanti, L., & Smith, S. M. (2014). Automatic denoising of functional MRI data: Combining independent component analysis and hierarchical fusion of classifiers. *NeuroImage*, *90*, 449–468. https://doi.org/10.1016/j.neuroimage.2013.11.046

Sanz Leon, P., Knock, S. A., Woodman, M. M., Domide, L., Mersmann, J., McIntosh, A. R., & Jirsa, V. (2013). The Virtual Brain: A simulator of primate brain network dynamics. *Frontiers in Neuroinformatics*, *7*. https://doi.org/10.3389/fninf.2013.00010

Schreiber, T. (2000). Measuring Information Transfer. *Physical Review Letters*, *85*(2), 461–464. https://doi.org/10.1103/PhysRevLett.85.461

Schultz, D. H., & Cole, M. W. (2016). Higher Intelligence Is Associated with Less Task-Related Brain Network Reconfiguration. *Journal of Neuroscience*, *36*(33), 8551–8561. https://doi.org/10.1523/JNEUROSCI.0358-16.2016

Seguin, C., Heuvel, M. P. van den, & Zalesky, A. (2018). Navigation of brain networks. *Proceedings of the National Academy of Sciences of the United States of America*, *115*(24), 6297–6302. https://doi.org/10.1073/pnas.1801351115

Seguin, C., Razi, A., & Zalesky, A. (2019). Send-receive communication asymmetry in brain networks: Inferring directionality of neural signalling from undirected structural connectomes. *BioRxiv*, 573071. https://doi.org/10.1101/573071

Shannon, C. E. (1948). A Mathematical Theory of Communication. *Bell System Technical Journal*, *27*(3), 379–423. https://doi.org/10.1002/j.1538-7305.1948.tb01338.x

Shine, J. M., Bissett, P. G., Bell, P. T., Koyejo, O., Balsters, J. H., Gorgolewski, K. J., Moodie, C. A., & Poldrack, R. A. (2016). The Dynamics of Functional Brain Networks: Integrated Network States during Cognitive Task Performance. *Neuron*, *92*(2), 544–554. https://doi.org/10.1016/j.neuron.2016.09.018

Shine, J. M., & Poldrack, R. A. (2018). Principles of dynamic network reconfiguration across diverse brain states. *NeuroImage*, *180*, 396–405. https://doi.org/10.1016/j.neuroimage.2017.08.010

Smith, R. E., Tournier, J.-D., Calamante, F., & Connelly, A. (2012). Anatomically-constrained tractography: Improved diffusion MRI streamlines tractography through effective use of anatomical information. *NeuroImage*, *62*(3), 1924–1938. https://doi.org/10.1016/j.neuroimage.2012.06.005

Smith, R. E., Tournier, J.-D., Calamante, F., & Connelly, A. (2013). SIFT: Spherical-deconvolution informed filtering of tractograms. *NeuroImage*, *67*, 298–312. https://doi.org/10.1016/j.neuroimage.2012.11.049

Smith, R. E., Tournier, J.-D., Calamante, F., & Connelly, A. (2015). SIFT2: Enabling dense quantitative assessment of brain white matter connectivity using streamlines tractography. *NeuroImage*, *119*, 338–351. https://doi.org/10.1016/j.neuroimage.2015.06.092

Smith, S. M., Beckmann, C. F., Andersson, J., Auerbach, E. J., Bijsterbosch, J., Douaud, G., Duff, E., Feinberg, D. A., Griffanti, L., Harms, M. P., Kelly, M., Laumann, T., Miller, K. L., Moeller, S., Petersen, S., Power, J., Salimi-Khorshidi, G., Snyder, A. Z., Vu, A. T., … Glasser, M. F. (2013). Resting-state fMRI in the Human Connectome Project. *NeuroImage*, *80*, 144–168. https://doi.org/10.1016/j.neuroimage.2013.05.039

Sneppen, K., Trusina, A., & Rosvall, M. (2005). Hide-and-seek on complex networks. *EPL (Europhysics Letters)*, *69*(5), 853. https://doi.org/10.1209/epl/i2004-10422-0

Sotiropoulos, S. N., Jbabdi, S., Xu, J., Andersson, J. L., Moeller, S., Auerbach, E. J., Glasser, M. F., Hernandez, M., Sapiro, G., Jenkinson, M., Feinberg, D. A., Yacoub, E., Lenglet, C., Van Essen, D. C., Ugurbil, K., & Behrens, T. E. J. (2013). Advances in diffusion MRI acquisition and processing in the Human Connectome Project. *NeuroImage*, *80*, 125–143. https://doi.org/10.1016/j.neuroimage.2013.05.057

Sporns, O. (2010). *Networks of the Brain*. MIT Press.

Sporns, O. (2013). Network attributes for segregation and integration in the human brain. *Current Opinion in Neurobiology*, *23*(2), 162–171. https://doi.org/10.1016/j.conb.2012.11.015

Stam, C. J. (2014). Modern network science of neurological disorders. *Nature Reviews. Neuroscience*, *15*(10), 683–695. https://doi.org/10.1038/nrn3801

Stejskal, E. O., & Tanner, J. E. (1965). Spin Diffusion Measurements: Spin Echoes in the Presence of a Time Dependent Field Gradient. *The Journal of Chemical Physics*, *42*(1), 288–292. https://doi.org/10.1063/1.1695690

ThriftBooks. (n.d.). *The Meme Machine book by Susan Blackmore*. ThriftBooks. Retrieved October 24, 2019, from https://www.thriftbooks.com/w/the-meme-machine_susan-j-blackmore/310230/

Timme, N. M., & Lapish, C. (2018). A Tutorial for Information Theory in Neuroscience. *ENeuro*, *5*(3). https://doi.org/10.1523/ENEURO.0052-18.2018

Tipnis, U., Amico, E., Ventresca, M., & Goñi, J. (2018). Modeling communication processes in the human connectome through cooperative learning. *IEEE Transactions on Network Science and Engineering*, 1–1. https://doi.org/10.1109/TNSE.2018.2878487

Tournier, J.-D., Calamante, F., & Connelly, A. (2012). MRtrix: Diffusion tractography in crossing fiber regions. *International Journal of Imaging Systems and Technology*, *22*(1), 53–66. https://doi.org/10.1002/ima.22005

Van Essen, D. C., Ugurbil, K., Auerbach, E., Barch, D., Behrens, T. E. J., Bucholz, R., Chang, A., Chen, L., Corbetta, M., Curtiss, S. W., Della Penna, S., Feinberg, D., Glasser, M. F., Harel, N., Heath, A. C., Larson-Prior, L., Marcus, D., Michalareas, G., Moeller, S., … WU-Minn HCP Consortium. (2012). The Human Connectome Project: A data acquisition perspective. *NeuroImage*, *62*(4), 2222–2231. https://doi.org/10.1016/j.neuroimage.2012.02.018

Van Essen, David C., Smith, S. M., Barch, D. M., Behrens, T. E. J., Yacoub, E., & Ugurbil, K. (2013). The WU-Minn Human Connectome Project: An overview. *NeuroImage*, *80*, 62–79. https://doi.org/10.1016/j.neuroimage.2013.05.041

Vann, S. D., Aggleton, J. P., & Maguire, E. A. (2009). What does the retrosplenial cortex do? *Nature Reviews Neuroscience*, *10*(11), 792–802. https://doi.org/10.1038/nrn2733

Wibral, M., Lizier, J. T., & Priesemann, V. (2015). Bits from Brains for Biologically Inspired Computing. *Frontiers in Robotics and AI*, *2*. https://doi.org/10.3389/frobt.2015.00005

Wibral, M., Vicente, R., & Lizier, J. T. (2014). *Directed Information Measures in Neuroscience*. Springer.

Worrell, J. C., Rumschlag, J., Betzel, R. F., Sporns, O., & Mišić, B. (2017). Optimized connectome architecture for sensory-motor integration. *Network Neuroscience*, *1*(4), 415–430. https://doi.org/10.1162/NETN_a_00022

Yeo, B. T. T., Krienen, F. M., Sepulcre, J., Sabuncu, M. R., Lashkari, D., Hollinshead, M., Roffman, J. L., Smoller, J. W., Zöllei, L., Polimeni, J. R., Fischl, B., Liu, H., & Buckner, R. L. (2011). The organization of the human cerebral cortex estimated by intrinsic functional connectivity. *Journal of Neurophysiology, 106*(3), 1125–1165. https://doi.org/10.1152/jn.00338.2011

# Supplementary Information

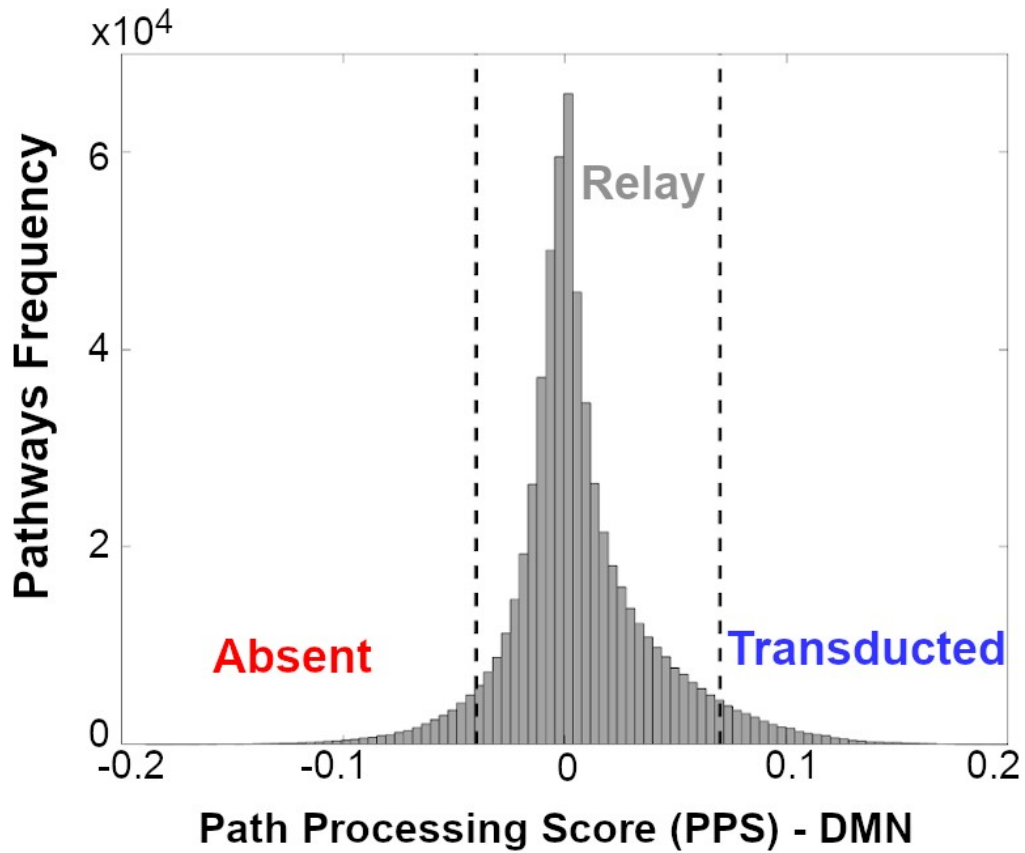

**Figure S1. Relay baseline for path processing score (PPS) communication regimes.** The DMN PPS distribution was obtained by considering the DMN-based shortest-paths and their corresponding (subject-level) PPS in resting state. Dashed vertical lines denote 5-95 percentiles respectively. Those percentiles were used as the range within which relay communications take place. Values below percentile 5 correspond to absent communication, whereas values above percentile 95 correspond to transducted communication.

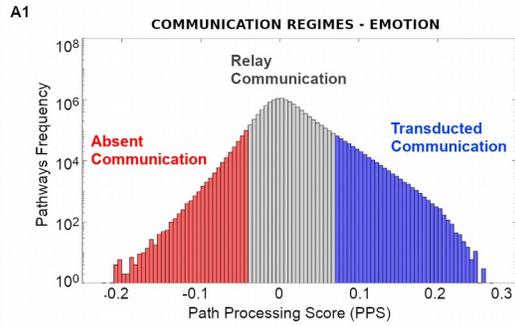
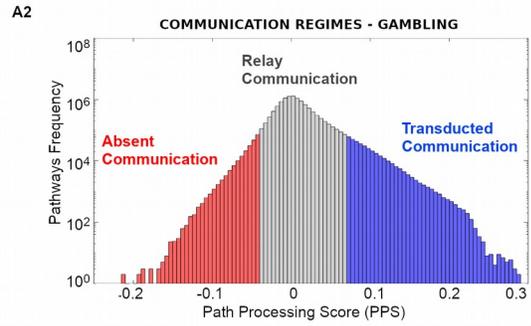
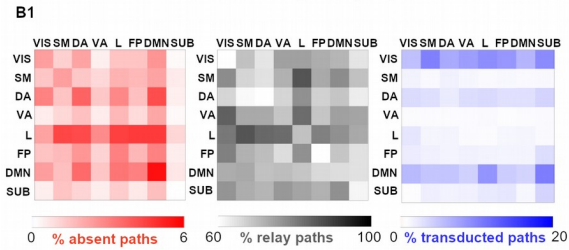
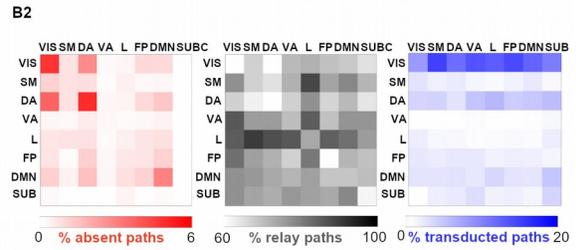
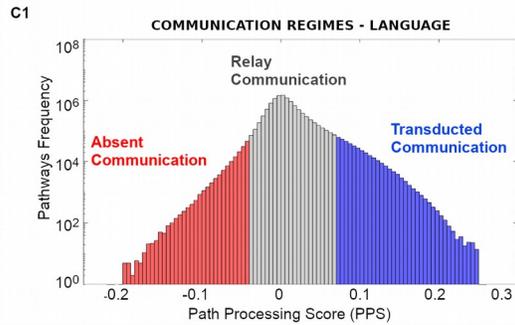
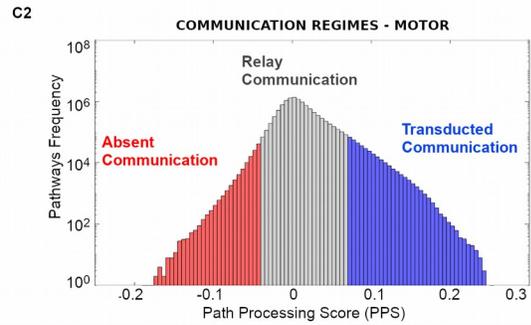
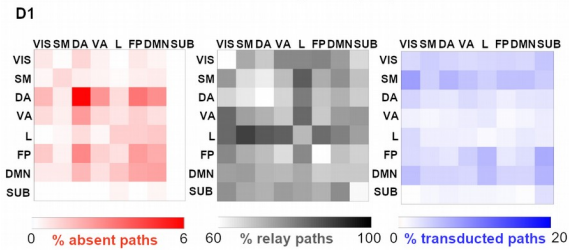
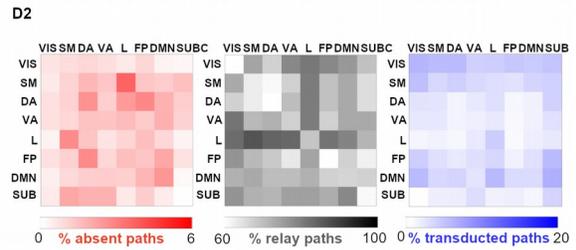
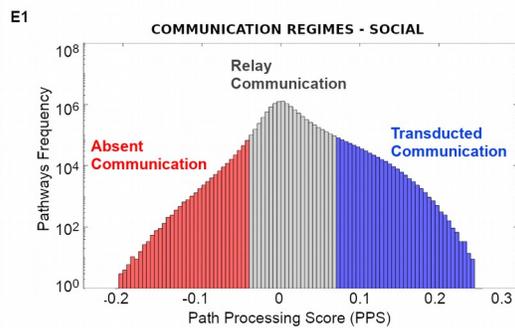
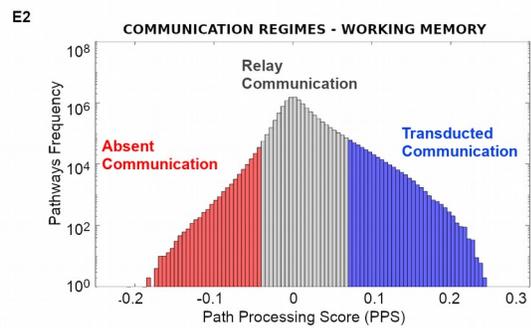
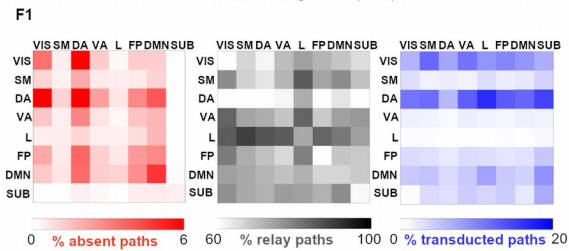
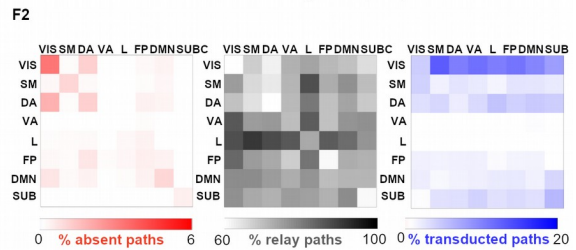

Figure S2. Communication regimes in large-scale brain networks. A1) Emotion task. Path Processing Score (PPS) on indirect pathways allows to separate brain network communication in three different regimes: absent, relay communication and transducted communication). B1) Emotion task. The percentage of paths, for the three different communication regimes, corresponding to the within and between 7 functional networks source-target pairs, as specified by (Yeo et al., 2011). An eighth sub-cortical community was added for completeness. Analogously, results are shown for **Gambling task (A2-B2)**, **Language task (C1-D1)**, **Motor task (C2-D2)**, **Social task (E1-F1)** and **Working Memory task (E2-F2)**.

|  | MI bin-width 0.5 | MI bin-width 0.75 | MI bin-width 1 | MI bin-width 2 |
|---|---|---|---|---|
| MI bin-width 0.5 | 1 | 0.98 | 0.95 | 0.87 |
| MI bin-width 0.75 | 0.98 | 1 | 0.98 | 0.90 |
| MI bin-width 1 | 0.95 | 0.98 | 1 | 0.94 |
| MI bin-width 2 | 0.87 | 0.90 | 0.93 | 1 |

Table S1. Effect of MI bin-width on MI-based functional connectomes. Table shows resting-state group average similarity between MI connectomes computed with different bin-widths (0.50, 0.75, 1 and 2, respectively). Note how the bin-width parameter minimally affects the MI connectome computation for the range of bin-widths explored.

|  | REST | EMOT | GAMB | LANG | MOT | RELAT | SOC | WM |
|---|---|---|---|---|---|---|---|---|
| **PPS r** | 0.98 | 0.96 | 0.96 | 0.95 | 0.93 | 0.97 | 0.97 | 0.97 |
| **PPS SD** | 0.017 | 0.021 | 0.019 | 0.017 | 0.018 | 0.021 | 0.019 | 0.016 |
| **PBS r** | 0.99 | 0.99 | 0.99 | 0.99 | 0.99 | 0.99 | 0.99 | 0.99 |
| **PBS SD** | 0.010 | 0.010 | 0.010 | 0.010 | 0.009 | 0.010 | 0.009 | 0.010 |

Table S2. Stability of PBS and PPS between runs and across subjects. Table reports the PPS and PBS group-average similarity across HCP trials (LR and RL, see Methods for details).